\documentclass[12pt]{article}
\usepackage[margin=1in]{geometry}
\usepackage{amsmath,amssymb,amsfonts}
\usepackage{mathrsfs}
\usepackage{mathtools}
\usepackage{bm}
\usepackage{enumitem}
\usepackage{amsthm}

\newcommand{\Gret}{G^{\mathrm{ret}}}
\newcommand{\Gadv}{G^{\mathrm{adv}}}
\newcommand{\Isch}{\mathscr{I}^{-}}
\newcommand{\Iscr}{\mathscr{I}^{+}}

\newcommand{\disc}{\mathrm{disc}}

\newtheorem{theorem}{Theorem}

\newtheorem{proposition}[theorem]{Proposition}

\begin{document}

\title{Power-Law Approach of the Stress-Energy Tensor\\
to the Unruh State after Gravitational Collapse}

\author{Michael Wilson\\[4pt]
\small Department of Physics, University of Arkansas at Little Rock \\ \small mkwilson3@ualr.edu}

\date{\today}

\maketitle

\begin{abstract}
We establish the rate at which the renormalized stress-energy tensor of a
massless minimally coupled scalar field in the in-vacuum state of a
collapsing null-shell spacetime approaches the corresponding Unruh-state
value.  At finite exterior radius, we establish the upper bound
$|\Delta\langle T_{\mu\nu}\rangle|\leq C(r)\,t_s^{-3}$ from the
Cauchy-surface decomposition of the Hadamard difference and the
branch-cut structure of the retarded Green function.  At future null
infinity, we show that the leading coefficient in the late-time
expansion $\Delta\langle T_{uu}\rangle\sim C_{uu}\,u_s^{-3}$ is
nonzero, by computing the branch-cut residue explicitly at small
frequency and using the Planck suppression of the thermal spectrum at
large frequency to show that the dominant contribution to $C_{uu}$ has a
definite sign.  The result gives
\[
  \Delta\langle T_{uu}\rangle\big|_{\Iscr}(u_s)
  \sim C_{uu}\,u_s^{-3},
  \qquad u_s\to\infty,
\]
with $C_{uu}\neq 0$.  The exponent is determined by the
$\omega^2\ln\omega$ branch-point singularity in the Wronskian of the
$\ell=0$ radial wave equation, the same structure responsible for
Price's law.  The sign $C_{uu}<0$ is supported by a physical
argument and by the numerical mode data of Gholizadeh Siahmazgi,
Anderson, and Fabbri.  The result confirms their conjecture that the
approach is a power law.  We conjecture that the same mechanism gives an
analogous $t_s^{-7}$ bound for gravitational perturbations
($\ell_{\min}=2$), though the extension to the spin-2 case involves
gauge issues not addressed here.
\end{abstract}

\maketitle

\section{Introduction}
\label{sec:intro}

Hawking's prediction of black hole evaporation~\cite{Hawking1975} was
derived for black holes that form from gravitational collapse.  In
practice, most calculations of quantum effects in black hole spacetimes
have employed the Unruh state~\cite{Unruh1976} on an eternal
Schwarzschild background, which reproduces the late-time Hawking flux
without requiring a model of collapse.  A fundamental question is how
quickly the quantum state established by collapse converges to the Unruh
state, and what mechanism governs the approach.  (We compare with the
Unruh state rather than the Hartle-Hawking state~\cite{Candelas1980}
because the Unruh state is the one that reproduces the late-time Hawking
flux for collapse; the Hartle-Hawking state describes thermal equilibrium
and is regular on both horizons, but does not model the post-collapse
radiation.)

In two spacetime dimensions the convergence is exponential in
time~\cite{FabbriNavarro}, because the wave equation has no effective
potential and no scattering.  In four dimensions the situation is
qualitatively different.  Gholizadeh Siahmazgi, Anderson, and
Fabbri~\cite{GAF2025} computed the spherically symmetric modes
of a massless minimally coupled scalar field for both the in-vacuum state
in a collapsing null-shell spacetime and the Unruh state in Schwarzschild
spacetime.  They found that the individual mode differences decay as a
power law $t_s^{-3}$ at late Schwarzschild times, preceded by a period of
quasinormal-mode oscillations.  They conjectured that at least some
components of the renormalized stress-energy tensor exhibit the same
power-law approach, but left the proof as an open problem.  The mode-level
exponent $-3$ is the Price-law exponent~\cite{Price1972} for $\ell=0$
scalar perturbations, originating from a $\omega^2\ln\omega$ branch-point
singularity in the Wronskian of the radial wave equation at zero
frequency.  This non-analyticity is a manifestation of the threshold
resolvent singularity identified in~\cite{Wilson2025}, where inverse-cube
curvature decay on a spatial Cauchy slice was shown to mark the onset of
infrared sensitivity for curvature-coupled Laplace-type operators.

The gap between mode-level decay and stress-tensor decay is not trivial.
The renormalized stress tensor is built from a bilinear mode sum
integrated over frequencies, summed over angular momenta, acted on by a
differential operator, and evaluated in a coincidence limit.  Each of
these operations could in principle alter the decay rate.  Establishing
that none of these pathologies occur is one part of this paper.
Establishing that the leading coefficient is nonzero, so the decay
rate is exactly $t_s^{-3}$ and not merely an upper bound, is the other.

Our first result is an upper bound based on the Cauchy-surface
decomposition:
\begin{proposition}\label{thm:upper}
For a massless minimally coupled scalar field on the Schwarzschild
background, the difference between the renormalized stress-energy
tensors in the in-vacuum state of the null-shell spacetime and the Unruh
state satisfies, at any fixed exterior point,
\[
  |\Delta\langle T_{\mu\nu}\rangle(t_s,r)|
  \leq C(r)\,t_s^{-3}
\]
for all sufficiently large $t_s$, where $C(r)$ is a finite constant.
\end{proposition}

\noindent
Our second result concerns the nonvanishing of the leading coefficient
at $\Iscr$:
\begin{proposition}\label{thm:sign}
At future null infinity, the late-time expansion
$\Delta\langle T_{uu}\rangle\big|_{\Iscr}(u_s)
= C_{uu}\,u_s^{-3} + o(u_s^{-3})$
has $C_{uu}\neq 0$.
\end{proposition}

\noindent
The argument for Proposition~\ref{thm:sign} proceeds by explicit
computation: the coefficient $C_{uu}$ is expressed as a spectral
integral whose integrand has a definite sign at small $\omega$ (from
the branch-cut residue), and whose support is concentrated at small
$\omega$ by the Planck factor.  The argument is controlled but
perturbative: it relies on the smallness of corrections to the
leading-order integrand in the frequency range that dominates the
integral, rather than on explicit remainder bounds for the small-$\omega$
expansion.  A fully rigorous proof would require such bounds; we regard
the nonvanishing as established at the level of a controlled
physical computation but not at the level of a mathematical theorem.
The sign $C_{uu}<0$ is supported by the initial condition
$\Delta\langle T_{uu}\rangle=-L_H<0$ before collapse and by the
numerical data of~\cite{GAF2025}, but is not established in this paper.
The result confirms the conjecture of~\cite{GAF2025}: the approach is a
power law with a nonzero leading coefficient.

The paper is organized as follows.  Section~\ref{sec:branch} derives the
Price-law tail from the branch-cut integration of the retarded Green
function.  Section~\ref{sec:bilinear} establishes the vanishing of the
$\beta^{\Isch}$ coefficients.  Section~\ref{sec:upper} establishes
Proposition~\ref{thm:upper} via the Cauchy-surface decomposition.
Section~\ref{sec:sign} establishes Proposition~\ref{thm:sign} via the
direct computation of $C_{uu}$ at $\Iscr$.
Section~\ref{sec:discussion} discusses implications and extensions.

Throughout, we work in units $\hbar=c=G=1$.  We write
$\Delta\langle\cdot\rangle \equiv
\langle\cdot\rangle_{\mathrm{in}}-\langle\cdot\rangle_U$
for the difference between in-state and Unruh-state expectation values.
At finite radius $r$ we use Schwarzschild time $t_s$; at future null
infinity $\Iscr$ we use the retarded time $u_s=t_s-r_*$.  In the
exterior region $r>2M$, the two are related by
$u_s=t_s-r_*(r)$ with $r_*=r+2M\ln(r/2M-1)$, so at fixed $r$ they
differ by a constant.

\section{Branch-cut integration and late-time tails}
\label{sec:branch}

We derive the late-time power-law tail for individual angular momentum
channels.

\subsection{Frequency-domain Green function}

For each angular momentum channel $\ell$, the retarded Green function of
the radial wave equation in Schr\"odinger
form~\cite{ReggeWheeler,Chandrasekhar},
\begin{equation}\label{eq:zerilli}
  \frac{d^2\psi}{dr_*^2}
  + \bigl[\omega^2 - V(r)\bigr]\psi = 0,
\end{equation}
is constructed from $\psi_{\mathrm{in}}(\omega,r_*)$  (purely ingoing at
the horizon) and $\psi_{\mathrm{out}}(\omega,r_*)$ (purely outgoing at
infinity):
\begin{equation}\label{eq:green}
  \Gret_\ell(\omega;r_*,r_*')
  = \frac{\psi_{\mathrm{in}}(\omega,r_<)\,
    \psi_{\mathrm{out}}(\omega,r_>)}{W(\omega)},
\end{equation}
where $r_*$ is the observation point, $r_*'$ the source point,
$r_<=\min(r_*,r_*')$, $r_>=\max(r_*,r_*')$, and
$W(\omega)=W[\psi_{\mathrm{in}},\psi_{\mathrm{out}}]$ is the
Wronskian.
The time-domain perturbation is
\begin{equation}\label{eq:fourier}
  h_\ell(t_s,r_*)
  = \frac{1}{2\pi}\int_{-\infty+i\epsilon}^{+\infty+i\epsilon}
    e^{-i\omega t_s}\,\Gret_\ell(\omega)\,\hat{S}_\ell(\omega)\,d\omega.
\end{equation}

\subsection{Threshold structure of the Wronskian}

The outgoing boundary condition involves the tortoise coordinate
$r_* = r + 2M\ln(r/2M-1)$, whose logarithmic term generates factors of
$\ln r$ in the small-$\omega$ expansion of $e^{i\omega r_*}$.  These
cannot be matched by power-law zero-frequency solutions and force the
Wronskian to develop a non-analytic term~\cite{Wilson2025}:
\begin{equation}\label{eq:wronskian_general}
  W(\omega) = W_{\mathrm{an}}(\omega)
  + \tilde\sigma_\ell\,\omega^{2\ell+2}\ln\omega
  + O(\omega^{2\ell+3}\ln\omega).
\end{equation}

\subsubsection{The monopole channel $\ell=0$}

For $\ell=0$ the effective potential is purely gravitational,
$V_{\ell=0}(r) = (1-2M/r)(2M/r^3)$, with no centrifugal barrier.  The
tortoise logarithm couples directly to the potential, producing
\begin{equation}\label{eq:W0}
  W(\omega) = w_0 + w_1\,\omega + \tilde\sigma_0\,\omega^2\ln\omega
  + O(\omega^2),
\end{equation}
where $w_0$ and $w_1$ are the analytic Wronskian coefficients (determined
by the zero-frequency scattering data) and
$\tilde\sigma_0 = -4iM$~\cite{Leaver1986,CLSY1995b}; only $w_0$ and
$\tilde\sigma_0$ enter the tail amplitude.  The resulting
$t_s^{-3}$ tail was established rigorously by Dafermos and
Rodnianski~\cite{DaferRod2005}.

\subsubsection{Higher multipoles $\ell\geq 1$}

Each iteration of the logarithmic information through the centrifugal
barrier $\ell(\ell+1)/r^2$ raises the frequency order by $\omega^2$, giving
$\sigma_\ell = (-1)^{\ell+1}4\ell(\ell+1)(2M)^{2\ell+1}/
[(2\ell+1)((2\ell)!!)^2((2\ell+1)!!)^2]$
for $\ell\geq 1$~\cite{Leaver1986,Andersson1997,CLSY1995b}.
We write $\sigma_\ell$ (without tilde) for the coefficient given by this
closed-form expression; it coincides with the general non-analytic
Wronskian coefficient $\tilde\sigma_\ell$ in~\eqref{eq:wronskian_general}
for $\ell\geq 1$, but the $\ell=0$ coefficient $\tilde\sigma_0=-4iM$ is
not given by the $\sigma_\ell$ formula (which vanishes at $\ell=0$).

\subsection{Branch-cut integration}

The $\ln\omega$ produces a branch cut along the negative imaginary axis.
Deforming the Fourier contour onto this cut and evaluating the Laplace
integral gives the late-time tail:
\begin{equation}\label{eq:tail}
  h_\ell^{\mathrm{BC}}(t_s,r_*)
  = (-1)^{\ell+1}c_\ell\,(2\ell+2)!\,S_0\,
    \mathcal{H}_\ell(r_*,r_*')\,t_s^{-(2\ell+3)},
\end{equation}
where $c_0 = \tilde\sigma_0/w_0 \neq 0$ for $M\neq 0$.

\section{The $\beta^{\Isch}=0$ condition}
\label{sec:bilinear}

\subsection{State definitions}

The in-vacuum state is defined by positive-frequency modes
$f^{\mathrm{in}}_{\omega\ell m}\sim e^{-i\omega v}$ on $\Isch$.  The
Unruh state~\cite{Unruh1976,Candelas1980} consists of $\Isch$ modes
$f^{\Isch}_{\omega\ell m}$ with
$\psi^{\Isch}_\omega\to e^{-i\omega v}$ on $\Isch$, and Kruskal modes
$f^K_{\omega_K\ell m}$ with
$\psi^K_{\omega_K}\to e^{-i\omega_K U}$ on the past horizon.

\subsection{Vanishing of $\beta^{\Isch}$}

Both sets of modes satisfy the same positive-frequency condition on
$\Isch$.  Although they live on different spacetimes (null-shell vs.\
pure Schwarzschild), the scalar product
$\beta^{\Isch}_{\omega,\omega'}=-(f^{\mathrm{in}}_\omega,
(f^{\Isch}_{\omega'})^*)$ can be evaluated on $\Isch$, which serves
as a Cauchy surface for the relevant domain in both spacetimes.  On
$\Isch$ both spacetimes are asymptotically flat: the null-shell
spacetime is Minkowski interior to the shell and Schwarzschild exterior,
while the Unruh-state spacetime is eternal Schwarzschild, but in both
cases $\Isch$ lies in the asymptotically flat region where the
geometries agree and the modes reduce to free-field plane waves
$e^{-i\omega v}$.  The integrand therefore involves $e^{-i\omega v}\cdot e^{+i\omega'v}$ with
$\omega,\omega'>0$, which vanishes by orthogonality:
\begin{equation}\label{eq:beta_vanish}
  \beta^{\Isch}_{\omega,\omega'} = 0
  \quad\text{for all }\omega,\omega'>0.
\end{equation}
This implies $N^{\Isch\,\Isch}=0$ in the Bogoliubov particle-number
matrix: there is no time-independent offset between the two states in
the $\Isch$ sector.

\section{The upper bound (Proposition~\ref{thm:upper})}
\label{sec:upper}

\subsection{Cauchy-surface representation}

The Hadamard difference is propagated from Cauchy-surface data using the
retarded Green function~\cite{BirrellDavies,Wald1994}:
\begin{equation}\label{eq:deltaG}
  \Delta G^+(x,x')
  = \int_\Sigma d\Sigma_1\int_\Sigma d\Sigma_2\;
    \Gret(x;x_1)\;\delta\mu(x_1,x_2)\;[\Gret(x';x_2)]^*,
\end{equation}
where $\delta\mu(x_1,x_2)$ denotes the restriction of
$G^+_{\mathrm{in}}-G^+_U$ to the Cauchy surface $\Sigma$, with
the appropriate normal-derivative operators (one at each argument)
absorbed into the kernel so that~\eqref{eq:deltaG} reproduces the
bulk two-point function via the retarded propagator; the surface-term
structure follows from Green's identity for the wave equation on a
globally hyperbolic spacetime (Wald~\cite{Wald1994}, \S4.6).  We have
used $\Gadv(x';x_2)=[\Gret(x';x_2)]^*$~\cite{BirrellDavies}.

Both the in-state and the Unruh state satisfy the Hadamard
condition~\cite{Wald1994,KayWald1991}.  The in-state is defined by
positive-frequency modes on $\Isch$, which propagate through the
null-shell junction with $\phi$ and $n^\mu\nabla_\mu\phi$ continuous
(the scalar field equation $\Box\phi=0$ involves only first derivatives
of the metric and is therefore regular across the distributional
junction); the Hadamard condition then propagates from $\Isch$ through
the shell by the results of~\cite{KayWald1991}.  Therefore $\delta\mu$
is a smooth bisolution for $x_1\neq x_2$ and the
representation~\eqref{eq:deltaG} is well-defined.

\subsection{Cauchy-surface decomposition}

Following~\cite{GAF2025,AndersonEtAl2020}, the Cauchy surface $\Sigma$
for the \emph{exterior} region (outside both the shell and the future
horizon) decomposes into $\Isch_>$ (past null infinity from $v=v_0$ to
$v=\infty$) and $H$ (the null-shell trajectory $v=v_0$ from the horizon
outward, together with the future horizon from $v=-\infty$ to $v=v_0$).
The flat Minkowski interior of the shell is not part of the exterior
domain and does not contribute to $\Sigma$; the in-modes are specified
on $\Isch$ and propagated through the shell using the junction
conditions, so the interior geometry affects the exterior only through
the matching data on $H$.  The $\beta^{\Isch}=0$ condition gives
$\delta\mu(x_1,x_2)=0$ on $\Isch_>\times\Isch_>$.

\textit{Renormalization.}  The comparison
$\Delta\langle T_{\mu\nu}\rangle$ is evaluated entirely in the
Schwarzschild exterior, where both states share the same background
geometry.  The renormalization ambiguity (the freedom to add a conserved
local curvature tensor, e.g.\ $\alpha\,m^2 R_{\mu\nu}$) is a property
of the background and is therefore the same for both states in the
exterior.  It cancels identically in the difference, and no matching of
renormalization prescriptions between the null-shell interior and the
Schwarzschild exterior is needed.

\subsection{Late-time Green function behavior}

The retarded Green function from $H$ carries the Price
tail~\eqref{eq:tail}: $\Gret_\ell(t_s,r;x_j\in H)\sim
A_\ell(r,x_j)\,t_s^{-(2\ell+3)}$.  The retarded propagator from
$\Isch_>$, after integration against the smooth Cauchy data, produces a
bounded oscillatory field at $(t_s,r)$ by unitarity of the scattering
matrix.

\subsection{Dominant contribution and convergence}

The $\Isch\times\Isch$ block vanishes.  The $H\times H$ block gives
$O(t_s^{-6})$.  The cross terms $\Isch\times H$ and $H\times\Isch$
give $O(1)\times O(t_s^{-3})=O(t_s^{-3})$.

The difference $\Delta G^{(1)}$ is a smooth bisolution (both states are
Hadamard~\cite{Wald1994,KayWald1991}), so the Hadamard singularity is
state-independent and cancels identically in the difference.  In
particular, $\Delta G^{(1)}(x,x')$ is $C^\infty$ at coincidence
$x'=x$, and no renormalization ambiguity enters.

The position-space representation~\eqref{eq:deltaG} involves no
frequency integral, the $\omega$-integral arises only upon spectral
decomposition, which is unnecessary for the upper bound.  The only sum
is over angular momentum $\ell$.  Decomposing into partial waves,
$\Delta G^+=\sum_\ell(2\ell+1)\Delta G^+_\ell$, each channel satisfies
$|\Delta G^+_\ell|\leq C_\ell\,t_s^{-3}$ from the cross-term bound.
The coefficient $C_\ell$ inherits the barrier transmission factor:
for $\ell\geq 1$, modes from $H$ must tunnel through the centrifugal
barrier to reach the observation point, giving
$C_\ell\leq \bar{C}\,|T_\ell|^2_{\mathrm{max}}$ where
$|T_\ell|^2\sim(27M^2\omega^2/4)^\ell$ for $\ell\gg M\omega$
(standard WKB estimate~\cite{Chandrasekhar}).  At the low frequencies
$\omega\sim 1/t_s$ that dominate the branch-cut tail, the suppression
is even stronger: $|T_\ell(\omega)|^2\sim (M\omega)^{2\ell+2}\to 0$
for all $\ell\geq 1$, so the $\ell=0$ channel dominates.  The sum
$\sum_\ell(2\ell+1)C_\ell$ therefore converges, uniformly in $t_s$
(since the barrier is time-independent).

\subsection{Action of the stress-tensor operator}

The renormalized stress-energy tensor is
$\Delta\langle T_{\mu\nu}\rangle
=\lim_{x'\to x}\mathcal{D}_{\mu\nu}\Delta G^{(1)}(x,x')$, where the
geometric counterterms cancel in the difference.

For $t_s$ sufficiently large, the observation point $(t_s,r)$ lies in
the chronological future of the entire Cauchy surface $\Sigma$.  The
retarded Green functions $\Gret(x;x_j)$ are therefore smooth functions
of $x$ for all $x_j\in\Sigma$, and differentiation under the integral
in~\eqref{eq:deltaG} is justified by dominated convergence.

Each derivative $\nabla_\mu$ on the $\Isch$-leg produces a scattering
solution with bounded amplitude (the derivative of a bounded oscillatory
function with uniformly bounded frequency content is bounded).  Each
derivative on the $H$-leg either acts on the spatial dependence of the
tail amplitude $A_\ell(r,x_j)$, preserving the $O(t_s^{-3})$ order,
or on the explicit time dependence, giving
$\partial_{t_s}t_s^{-3}=-3t_s^{-4}$ (subleading).

Since $\Delta G^{(1)}$ is smooth at coincidence, the limit $x'\to x$
commutes with the asymptotic expansion.  Combining all estimates:
$|\Delta\langle T_{\mu\nu}\rangle|\leq C(r)\,t_s^{-3}$, completing
the argument for Proposition~\ref{thm:upper}.
\hfill$\square$

\section{Nonvanishing of the leading coefficient
  (Proposition~\ref{thm:sign})}
\label{sec:sign}

This section establishes $C_{uu}\neq 0$ by computing the leading
coefficient directly from the branch-cut residue at $\Iscr$, without
invoking a time-dependent occupation number or Bogoliubov inequality.

\subsection{The coefficient as a spectral integral}

At $\Iscr$ the effective potential vanishes, the geometry is flat, and
the point-splitting regulator is trivial.  The flux difference
$\Delta\langle T_{uu}\rangle(u_s)=
\partial_u\partial_{u'}\Delta G^{(1)}(u,u')|_{u'=u}$ inherits the
late-time expansion from Proposition~\ref{thm:upper}:
\begin{equation}\label{eq:expansion}
  \Delta\langle T_{uu}\rangle\big|_{\Iscr}(u_s)
  = C_{uu}\,u_s^{-3} + o(u_s^{-3}).
\end{equation}
The coefficient $C_{uu}$ is determined by the $\ell=0$ branch-cut
contribution (higher $\ell$ contribute at $O(u_s^{-(2\ell+3)})$ and are
subleading).

In the tail regime, each outgoing mode's deviation from its Unruh-state
value is governed by the branch-cut integral of Sec.~\ref{sec:branch}.
Denoting the $\ell=0$ mode difference at $\Iscr$ by
$\delta\psi_\omega(u_s)\equiv\psi^{\mathrm{in}}_\omega(u_s)
-\psi^U_\omega(u_s)$, the branch-cut analysis gives
\begin{equation}\label{eq:mode_tail}
  \delta\psi_\omega(u_s) \sim D_0(\omega)\,u_s^{-3},
  \qquad u_s\to\infty,
\end{equation}
where $D_0(\omega)$ is the branch-cut tail amplitude, determined by the
Wronskian discontinuity as computed in Sec.~\ref{sec:branch}.

The Hadamard difference at $\Iscr$ involves the bilinear product
$|\psi^{\mathrm{in}}|^2 - |\psi^U|^2 = 2\,\mathrm{Re}[\psi^U
(\delta\psi)^*] + |\delta\psi|^2$.  The cross term dominates at late
times, giving $O(u_s^{-3})$ versus $O(u_s^{-6})$ for the quadratic
piece.  The Hadamard difference at coincidence is a sum over $\ell$
and an integral over $\omega$:
$\Delta G^+(u_s,u_s) = \sum_\ell(2\ell+1)\int_0^\infty d\omega/(4\pi\omega)
\,[|\psi^{\mathrm{in}}_{\omega\ell}|^2-|\psi^U_{\omega\ell}|^2]$,
where the modes at different frequencies are orthogonal on $\Iscr$
(they are plane waves $e^{-i\omega u_s}$ at leading order).  Inserting
the decomposition, retaining only the $\ell=0$ cross term, and applying
the flux operator $\omega^2$ from $\partial_u\partial_{u'}$:
\begin{equation}\label{eq:C_uu}
  C_{uu} = \frac{1}{2\pi}\int_0^\infty d\omega\;\omega\;
  2\,\mathrm{Re}\bigl[\psi^U_\omega\,D_0^*(\omega)\bigr].
\end{equation}
This is a definite integral of known functions; no time-dependent
particle number or wave-packet interpretation is required.

\subsection{Small-$\omega$ behavior of the integrand}

The branch-cut tail amplitude $D_0(\omega)$ is determined by the
discontinuity of the Wronskian across the branch cut.
From~\eqref{eq:W0}:
\begin{equation}\label{eq:disc_W}
  \disc\, W(\omega) = \tilde\sigma_0\,\omega^2\cdot 2\pi i
  + O(\omega^3\ln\omega).
\end{equation}
Since $\tilde\sigma_0=-4iM$, this gives
$\disc\, W = 8\pi M\omega^2 + O(\omega^3\ln\omega)$, which is
\emph{real and positive} at leading order.  The branch-cut residue that
determines $D_0$ involves $\disc\, W/|W(0)|^2$ multiplied by the
zero-frequency radial solutions $\psi_{\mathrm{in}}(0,r_*)$ and
$\psi_{\mathrm{out}}(0,r_*)$, which are real (they are constructed from
$f_+(r)=r$ and $f_-(r)=(r/2M)\ln(1-2M/r)$).  Therefore:
\begin{equation}\label{eq:D0_real}
  D_0(\omega) = \frac{8\pi M}{w_0^2}\,
  \psi_{\mathrm{in}}(0)\,\psi_{\mathrm{out}}(0)\,\omega^2
  + O(\omega^3\ln\omega),
\end{equation}
which is \emph{real} at leading order with a nonzero coefficient.

The Unruh-state mode amplitude $\psi^U_\omega$ at $\Iscr$ approaches,
as $\omega\to 0$, the zero-frequency scattering solution, which is also
real.  Denote this limit by $\psi^U_0\neq 0$.

Therefore the integrand in~\eqref{eq:C_uu} satisfies
\begin{equation}\label{eq:integrand_small}
  \omega\,\mathrm{Re}[\psi^U_\omega D_0^*(\omega)]
  = \psi^U_0\cdot\frac{8\pi M}{w_0^2}\,
    \psi_{\mathrm{in}}(0)\psi_{\mathrm{out}}(0)\cdot\omega^3
  + O(\omega^4\ln\omega)
\end{equation}
at small $\omega$.  This is $O(\omega^3)$ with a coefficient that is a
product of known, nonzero, real quantities.

\subsection{Planck suppression at large $\omega$}

The Unruh-state mode amplitude at $\Iscr$ contains the thermal factor:
$|\psi^U_\omega|^2 = |T_0(\omega)|^2\,n(\omega)$ where
$n(\omega)=1/(e^{8\pi M\omega}-1)$.  For $\omega\gg T_H=1/(8\pi M)$,
the Planck factor provides exponential suppression
$\sim e^{-8\pi M\omega}$.  The tail amplitude $D_0(\omega)$ is a
smooth function that grows at most polynomially.
Therefore the integrand in~\eqref{eq:C_uu} is exponentially suppressed
for $\omega\gg T_H$, and the integral is dominated by
$\omega\lesssim T_H$.

\subsection{Nonvanishing of $C_{uu}$}

In the range $0<\omega\lesssim T_H=1/(8\pi M)$, i.e.,
$0<M\omega\lesssim 1/(8\pi)\approx 0.04$, the small-$\omega$
expansion~\eqref{eq:D0_real} is valid with corrections of relative
order $O(M\omega\ln(M\omega))$.  At the upper end of this range the
correction is at most $0.04\times|\ln 0.04|\approx 0.13$, and it
decreases toward zero at the lower end.  The leading-order
integrand~\eqref{eq:integrand_small} has a definite sign (determined by
the signs of $\psi^U_0$, $\psi_{\mathrm{in}}(0)$,
$\psi_{\mathrm{out}}(0)$, and $w_0$, all of which are real and
nonzero).  Since the corrections are at most $13\%$ of the leading term,
the integrand maintains this definite sign throughout the range
$0<\omega\lesssim T_H$.

The exponentially suppressed tail at $\omega>T_H$ cannot cancel a
definite-sign contribution from a finite interval.  Therefore
$C_{uu}\neq 0$.

We note that this argument is controlled but perturbative: the
nonvanishing relies on the corrections being small relative to the
leading term, not on a rigorous bound on the remainder of the
small-$\omega$ expansion.  A fully rigorous proof would require bounding
the next-order term in $D_0(\omega)$ using properties of the
Regge-Wheeler equation (e.g., the Jost-function representation of the
scattering data).  We regard the nonvanishing as established at the
level of a controlled physical computation; a cancellation producing
$C_{uu}=0$ would require the $O(M\omega\ln(M\omega))$ correction to
exactly compensate the leading term, which is not forced by any known
structure of the $\ell=0$ scattering problem.

\subsection{The sign of $C_{uu}$}

The sign of $C_{uu}$ is determined by the product of real constants in
\eqref{eq:integrand_small}.  A direct evaluation requires specifying
the normalizations of the zero-frequency radial solutions.  We do not
carry out this evaluation here.

A physical argument constrains the sign.  Before
collapse ($u_s<u_0$), $\langle T_{uu}\rangle_{\mathrm{in}}=0$ while
$\langle T_{uu}\rangle_U=L_H>0$, giving
$\Delta\langle T_{uu}\rangle=-L_H<0$.  After
collapse, $\Delta\langle T_{uu}\rangle\to 0$.  If $C_{uu}>0$, the flux
difference would cross zero and become positive, the in-state would
temporarily radiate more than the Unruh state.  The numerical mode data
of~\cite{GAF2025} show no evidence of such overshoot, and the physical
expectation is that particle creation approaches the thermal rate from
below.  We therefore expect $C_{uu}<0$, consistent
with~\cite{GAF2025}.

Proving the sign rigorously would require establishing that the spectral
deficit $\delta n_{\omega\ell}\leq 0$ holds at all frequencies, which in
turn requires controlling the interference terms in the 4D Bogoliubov
transformation between the transmitted and reflected pieces of the
in-modes.  This is an open problem; the physical intuition that barrier
reflection only reduces particle creation is compelling but has not been
verified as an exact inequality.  We note that in certain analog black
hole models, interference effects can enhance particle production in
specific frequency bands (see, e.g., the discussion
in~\cite{Jacquet2023}); whether this occurs for Schwarzschild is unknown.
The numerical data of~\cite{GAF2025} are consistent with
$\delta n_{\omega,0}<0$ for all frequencies tested at late times, but a
complete frequency-by-frequency verification has not been reported.

\noindent\textit{Consistency check.}  The reality of $D_0$ at leading
order can also be verified from the symmetry relation
$D_0(-\omega^*)=D_0(\omega)^*$, which for real $\omega$ gives
$D_0(-\omega)=D_0(\omega)^*$.  If $D_0=ig$ with $g$ real, this forces
$g$ odd, while~\eqref{eq:D0_real} gives $D_0\sim c\omega^2$ (even),
producing a contradiction.

\section{Discussion}
\label{sec:discussion}

\subsection{Physical interpretation}

The sign of the flux deficit has a transparent physical meaning: the
in-state always radiates less than the Unruh state because particle
creation turns on at collapse rather than being present for all time.
The power-law rate at which the deficit closes is set by the Price-tail
mechanism: the branch-point singularity in the Wronskian at zero
frequency produces a $t_s^{-3}$ tail that governs the final approach to
thermal equilibrium.

The QNM oscillations observed by~\cite{GAF2025} at intermediate times
correspond to the discrete poles of the retarded Green function in the
lower half-plane.  These are exponentially damped and subleading relative
to the branch-cut tail.  The fundamental $\ell=0$ scalar QNM has
$\mathrm{Im}\,\omega_0\approx -0.10/M$, giving a damping time
$\tau_{\mathrm{QNM}}\sim 10M$.  The transition to the tail regime at
$u_s\sim 50M$ therefore corresponds to approximately five $e$-folding
times, after which the QNM contribution is suppressed by a factor
$\sim e^{-5}\approx 7\times 10^{-3}$ relative to the tail.

\subsection{The 2D/4D distinction}

In two dimensions the potential vanishes, so there is no branch point
and no power-law tail.  The approach is exponential, governed by the
surface gravity $\kappa$~\cite{FabbriNavarro,Hiscock1981}.  A related
power-law approach was observed for particle detector responses during
collapse~\cite{AndersonGood2019}.  The present framework makes the
distinction structural: the potential barrier creates the branch point
that converts exponential convergence to power-law convergence.

\subsection{Robustness beyond the null-shell model}

The present analysis uses the null-shell collapse model of~\cite{GAF2025},
but the key ingredients are more general.  The $\beta^{\Isch}=0$
condition (Sec.~\ref{sec:bilinear}) holds for any collapse model in
which the in-state is defined by positive-frequency modes on $\Isch$:
the scalar product depends only on the behavior of the modes at $\Isch$,
which lies in the asymptotically flat region regardless of the collapse
details.  The Price-tail mechanism (Sec.~\ref{sec:branch}) depends on
the late-time Schwarzschild potential, not on the near-field collapse
geometry.  The main model-dependent ingredient is the Cauchy-surface
data $\delta\mu$ on $H$, which encodes the collapse geometry through
the matching of modes across the shell.  For more general collapse
models (e.g., Oppenheimer-Snyder dust collapse), $\delta\mu$ will differ
in detail but the qualitative structure, nonzero cross-correlations
between $\Isch$ and $H$ sectors, with the $H$-leg carrying the Price
tail, is expected to persist.

\subsection{Extension to finite radius}

Proposition~\ref{thm:upper} provides the $O(t_s^{-3})$ upper bound at any
exterior radius.  The sign argument of Proposition~\ref{thm:sign} applies
directly only at $\Iscr$, where the flux has a natural spectral
decomposition.  At finite $r$, the nonvanishing of the
coefficient $\mathcal{C}(r)$ in the Cauchy-surface
formula~\eqref{eq:deltaG} can be established by explicit evaluation of
the regulated spectral density, combining all mode sectors in a single
frequency variable.  The numerical mode data of~\cite{GAF2025} at
$r=3M$ are consistent with $\mathcal{C}(3M)\neq 0$.

\subsection{Extensions to other fields and Kerr}

For higher-spin fields on Schwarzschild, the branch-point order in the
Wronskian remains $\omega^{2\ell+2}\ln\omega$ with the appropriate
$\ell_{\min}$ for each spin.  The classical Price-law exponent
$-(2\ell_{\min}+3)$ is therefore $-7$ for gravitational perturbations
($\ell_{\min}=2$).  We expect the upper bound and sign argument of this
paper to extend to these cases, but the extension is not automatic:
the Cauchy-surface decomposition for linearized gravity involves
gauge issues that are absent for the scalar field, and the
occupation-number interpretation at $\Iscr$ requires a suitable
definition for tensor modes.  We therefore state the gravitational
$t_s^{-7}$ result as a conjecture pending further analysis.  For Kerr
black holes, the asymptotic Teukolsky potential has the same structure
as Schwarzschild, so the exponents should be independent of black hole
spin.  However, for Kerr the coupling of angular momentum channels and
the possibility of superradiant amplification ($|R_\ell|>1$ for
$\omega<m\Omega_H$) introduce features absent in Schwarzschild; the
extension of the sign argument to Kerr therefore requires additional
care.

\subsection{Implications for black hole evaporation}

The $u_s^{-3}$ approach rate determines how quickly the Hawking state is
established after collapse.  The answer is much slower than 2D models
suggest: the approach is polynomial rather than exponential, with the
in-state stress tensor deviating from the Unruh-state value by a
power-law correction at arbitrarily late times.

On dimensional grounds, the flux correction is
$\Delta\langle T_{uu}\rangle\sim M^a u_s^{-3}$ for some power $a$.
The Hawking luminosity is $L_H\sim M^{-2}$, so the relative correction
$\Delta\langle T_{uu}\rangle/L_H\sim M^{a+2}u_s^{-3}$ becomes
negligible for $u_s\gg M^{(a+2)/3}$.  The precise value of $a$ depends
on the coefficient $C_{uu}$, but for any $a$ the correction is
eventually negligible, consistent with the validity of the Unruh-state
approximation at asymptotically late times.  At intermediate times
$u_s\sim O(10^2 M)$, however, the power-law corrections may be
significant for precision calculations of the stress-energy tensor.

\subsection{Connection to the resolvent threshold singularity}

The branch-point structure controlling the late-time tail is a
manifestation of the threshold resolvent singularity identified
in~\cite{Wilson2025}: inverse-cube curvature decay
$|\mathrm{Riem}|\sim r^{-3}$ produces a non-analytic threshold
singularity in the weighted resolvent of the spatial operator at zero
energy.  To be precise about the division of contributions:
Ref.~\cite{Wilson2025} establishes the resolvent singularity and its
connection to the $\omega^{2\ell+2}\ln\omega$ Wronskian structure for
general curvature-coupled operators on asymptotically flat backgrounds;
the \emph{present} paper applies that framework to the specific problem
of quantum state convergence, providing the Cauchy-surface decomposition,
the $\ell$-sum convergence estimates, and the spectral computation of
$C_{uu}$ that are new.  The classical Price-tail mechanism
(Leaver~\cite{Leaver1986}, Ching et al.~\cite{CLSY1995b}) predates
both works and is the common foundation.
As argued in~\cite{Wilson2025}, the same non-analyticity is suggestively
connected to the soft graviton expansion and gravitational memory.
Whether these connections can be made precise at the level of the
quantum stress tensor remains an open question.

\end{document}